\documentclass{article}
\usepackage{spconf,amsmath,graphicx}
\usepackage{hyperref}
\usepackage{caption}
\usepackage{algorithm,algorithmicx,algpseudocode}
\title{Minimally-Supervised Speech Synthesis with Conditional Diffusion Model and Language Model: A Comparative Study of Semantic Coding}

\name{Chunyu Qiang$^{1,2,3}$, Hao Li$^{2}$, Hao Ni$^{2}$, He Qu$^{2}$, Ruibo Fu$^{4}$, Tao Wang$^{4}$, Longbiao Wang$^{1,3,*}$, Jianwu Dang$^{3}$}

\address{$^1$School of New Media and Communication, Tianjin University, Tianjin, China \\
$^2$Kuaishou Technology Co., Ltd, Beijing, China \\
$^3$Tianjin Key Laboratory of Cognitive Computing and Application, \\
College of Intelligence and Computing, Tianjin University, Tianjin, China \\
$^4$Institute of Automation, Chinese Academy of Sciences, Beijing, China }

\begin{document}
\ninept
\maketitle

\begin{abstract}
Recently, there has been a growing interest in text-to-speech (TTS) methods that can be trained with minimal supervision by combining two types of discrete speech representations and using two sequence-to-sequence tasks to decouple TTS. However, existing methods suffer from three problems: the high-frequency waveform distortion of discrete speech representations, the prosodic averaging problem caused by the duration prediction model in non-autoregressive frameworks, and difficulty in prediction due to the information redundancy and dimension explosion of existing semantic coding methods. To address these problems, three progressive methods are proposed. First, we propose Diff-LM-Speech, an autoregressive structure consisting of a language model and diffusion models, which models the semantic embedding into the mel-spectrogram based on a diffusion model to achieve higher audio quality. We also introduce a prompt encoder structure based on a variational autoencoder and a prosody bottleneck to improve prompt representation ability. Second, we propose Tetra-Diff-Speech, a non-autoregressive structure consisting of four diffusion model-based modules that design a duration diffusion model to achieve diverse prosodic expressions. Finally, we propose Tri-Diff-Speech, a non-autoregressive structure consisting of three diffusion model-based modules that verify the non-necessity of existing semantic coding models and achieve the best results. Experimental results show that our proposed methods outperform baseline methods. We provide a website with audio samples. \href{https://qiangchunyu.github.io/Diff-LM-Speech/}{$^1$}

\end{abstract}

\renewcommand{\thefootnote}{\fnsymbol{footnote}} %将脚注符号设置为fnsymbol类型,即特殊符号表示
\footnotetext{$*$ Corresponding author.}
\footnotetext{Audio samples: https://qiangchunyu.github.io/Diff-LM-Speech/}
% \footnotetext{Preprint. Work in progress.}


\begin{figure*}[t]
 \centering
 \includegraphics[width=\linewidth]{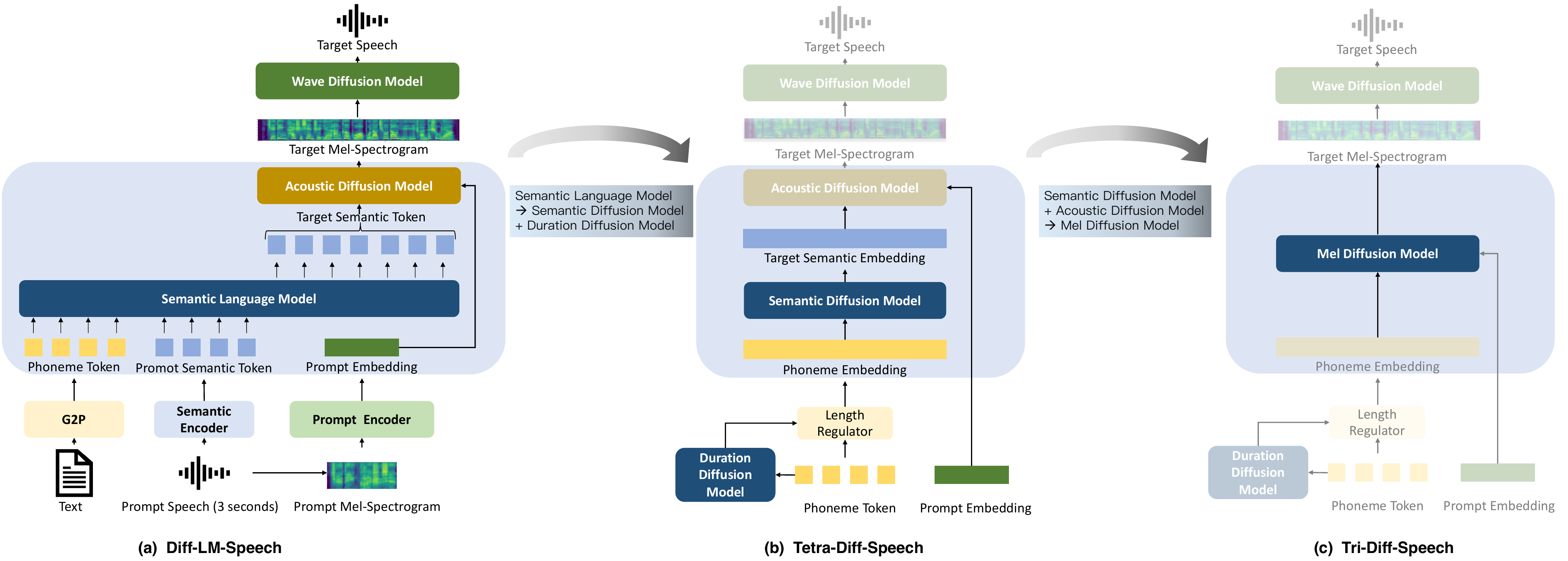}
 \vspace{15pt}
 \captionsetup{belowskip=15pt}
 \caption{The architecture of proposed models.}
 \label{fig:proposed_model}
\end{figure*}

\section{Introduction}
\label{sec:intro}
As deep learning advances, speech synthesis technology has made significant progress. Traditional speech synthesis methods have achieved satisfactory results\cite{wang2017tacotron,arik2017deep, li2019neural,ren2019fastspeech, kim2020glow, elias2021parallel}. The emergence of technologies such as GPT \cite{radford2018improving, brown2020language} has increased interest in large-scale TTS systems. These TTS systems can be broadly divided into two categories: 1) autoregressive frameworks \cite{borsos2022audiolm, wang2023neural,zhang2023speak,kharitonov2023speak} and 2) non-autoregressive frameworks \cite{levkovitch2022zero, shen2023naturalspeech, le2023voicebox}. 

Traditional speech synthesis methods typically use mel-spectrogram as intermediate representations. However, recent advancements in neural codec for speech \cite{baevski2020wav2vec, Hsu2021HuBERTSS, Defossez2022HighFN, Zeghidour2022SoundStreamAE} have led TTS methods to convert audio waveforms into discrete codes as intermediate representations. Notable examples include VALL-E \cite{wang2023neural}, the first large-scale TTS framework based on a language model with in-context learning capabilities for zero-shot speech synthesis. However, discrete acoustic coding relies on neural codecs for speech waveform reconstruction and suffers from information loss on high-frequency fine-grained acoustic details compared to traditional audio features. Additionally, the autoregressive framework suffers from the typical problems of instability and uncontrollability. Naturalspeech2 \cite{shen2023naturalspeech} is a non-autoregressive TTS framework based on a latent diffusion model \cite{ho2020denoising}. However, the duration prediction model required by non-autoregressive frameworks can cause expression averaging issues. SPEAR-TTS \cite{zhang2023speak} is another example that splits the TTS task into two tasks (text-to-semantic and semantic-to-speech) to achieve minimally-supervised training. The information content of the semantic coding is expected to be a "bridge" between text and acoustic information. It should emphasize linguistic content while de-emphasizing paralinguistic information such as speaker identity and acoustic details. However the semantic coding extracted by existing models suffers from excessive redundancy and dimension explosion, leading to difficulties in prediction from text and cumulative errors. To address these three issues, we propose three progressive methods:

1) Diff-LM-Speech, an autoregressive structure consisting of {\bf diff}usion models and a {\bf l}anguage {\bf m}odel, which models the semantic embedding into the mel-spectrogram based on a diffusion model to address the high-frequency waveform distortion issues of existing autoregressive methods based on language models. We also introduce a prompt encoder structure based on a variational autoencoder and a prosody bottleneck to improve prompt representation ability. 

2) Tetra-Diff-Speech, a non-autoregressive structure consisting of {\bf four diff}usion model-based modules that replace the semantic language model in Diff-LM-Speech with a semantic diffusion model. The duration diffusion model achieves diverse prosodic expressions and solves the expression averaging problem caused by the duration prediction model in non-autoregressive frameworks, as well as common problems of missing and repeated words in autoregressive frameworks.

3) Tri-Diff-Speech, a non-autoregressive structure consisting of {\bf three diff}usion model-based modules that compress and merge the semantic diffusion model and acoustic diffusion model in Tetra-Diff-Speech into a mel diffusion model. This structure verifies the non-necessity of semantic coding and avoids the problems of cumulative errors, information redundancy, and dimension explosion in existing semantic coding models.

\section{Method}

\subsection{Overview}

\subsubsection{Diff-LM-Speech}
\label{sec:diff-lm}
Diff-LM-Speech extends SPEAR-TTS\cite{zhang2023speak} by enabling the diffusion model for continuous-valued acoustic feature regression tasks. The framework has three main stages, as shown in Fig. \ref{fig:proposed_model} (a). In the first stage, the semantic language model translates text input into a sequence of discrete semantic tokens. The second stage maps the semantic embedding (continuous values from the codebook) into the mel-spectrogram by the acoustic diffusion model. The third stage maps the mel-spectrogram into the waveform by the wave diffusion model. Diff-LM-Speech performs two primary tasks: an autoregressive discrete coding classification task (text into semantic coding) and a non-autoregressive continuous valued prediction task (semantic coding into speech).

Diff-LM-Speech differs from VALL-E in several aspects: 1) Diff-LM-Speech is a three-stage model based on semantic coding, providing minimal supervised training capability that VALL-E lacks, making it different in model framework. 2) It uses mel-spectrogram as acoustic features, addressing high-frequency waveform distortion issues caused by discrete acoustic coding used in VALL-E and similar methods. 3) The use of acoustic diffusion model and wave diffusion model leads to improved speech quality. 4) Our designed prompt encoder enhances prompt representation ability.

\subsubsection{Tetra-Diff-Speech}
\label{sec:tetra-diff}
Tetra-Diff-Speech, as shown in Fig. \ref{fig:proposed_model} (b), consists of four diffusion-based modules that differ from Diff-LM-Speech. A non-autoregressive semantic diffusion model replaces the semantic language model in Diff-LM-Speech, achieving the prediction from text to semantic embedding. A duration diffusion model is designed to predict the corresponding duration of phonemes to solve the problem of mismatch between the length of phoneme sequences and semantic sequences. During training, the ground-truth duration is used to expand the phoneme sequence, while during inference, the corresponding predicted duration is used.

Tetra-Diff-Speech differs from NaturalSpeech2 in several aspects: 1) Tetra-Diff-Speech evolves from Diff-LM-Speech and is a three-stage model based on semantic coding, providing minimal supervised training capability that NaturalSpeech2 lacks, making it different in model framework. 2) It uses mel-spectrogram as acoustic features and includes acoustic diffusion model and wave diffusion model, which are not present in NaturalSpeech2. 3) Particularly, the proposed duration diffusion model addresses common prosody averaging issues in non-autoregressive structures like NaturalSpeech2.

\subsubsection{Tri-Diff-Speech}
\label{sec:tri-diff}

Tri-Diff-Speech, as shown in Fig. \ref{fig:proposed_model} (c), consists of three diffusion-based modules that use a mel diffusion model to predict mel-spectrogram directly from text. This framework aims to verify whether the two-stage process based on semantic coding in the existing semantic coding models is really effective relative to the traditional one-stage process. 

Tri-Diff-Speech differs from Voicebox \cite{le2023voicebox} in several aspects: 1) Tri-Diff-Speech evolves from Tetra-Diff-Speech and achieves direct prediction from text to mel-spectrogram using diffusion models. Voicebox is based on flow matching models. 2) Additionally, Tri-Diff-Speech includes duration diffusion model, prompt encoder, mel diffusion model, and wave diffusion model, which are not present in Voicebox.

\subsection{Prompt Feature Extractor}
\subsubsection{Semantic Encoder}

Similar to SPEAR-TTS\cite{zhang2023speak}, we use semantic coding as an intermediate representation between text and acoustic coding. To preserve high-frequency fine-grained acoustic details, we replace SoundStream's\cite{zeghidour2021soundstream} discrete acoustic coding with mel-spectrogram. The purpose of semantic coding is to provide coarse, high-level conditioning to subsequently produce the mel-spectrogram.  Thus, semantic coding should provide a representation of speech in which linguistic content (phonetics-to-semantics) is emphasized, while paralinguistic information such as speaker identity and acoustic details are de-emphasized. We obtain a 512-dimensional embedding with 1024 discrete values by fine-tuning a HuBert\cite{hsu2021hubert} model on an automatic speech recognition (ASR) task. This approach enables the model to learn a more robust and discriminative representation of speech that captures phonetic and semantic information. Additionally, we use Whisper's\cite{radford2023robust} encoder module as a control group to extract semantic coding (512-dimensional).

\subsubsection{Prompt Encoder}

The prompt encoder is a VAE-based model \cite{qiang2023improving} that extracts paralinguistic information, such as timbre, style, and prosody, from the prompt speech. It comprises a 6-layer 2D convolutional network and a SE-ResNet block \cite{hu2018squeeze}, which recalibrates channel-wise feature responses by modeling interdependencies among channels, resulting in significant performance improvements. The VAE structure enables the model to obtain a continuous and complete latent space distribution of styles, improving the ability to extract paralinguistic information. A 64-dimensional vector is sampled from the Gaussian distribution as the prompt embedding. To address the KL collapse problem, three tricks are used: 1) introducing KL annealing, 2) adopting a staged optimization method to optimize the reconstruction loss first and then the KL loss, and 3) A margin $\Delta$ is introduced to limit the minimum value of the KL loss as shown:

\begin{equation}
\begin{aligned}
\mathcal{L}_{kl} = max(0, D_{KL}[\mathcal{N}({\hat{\mu}},{\hat{\sigma}}^2)||\mathcal{N}(0, I)]-\Delta
\end{aligned}
\end{equation}

\begin{figure}[t]
 \centering
 \includegraphics[width=\linewidth]{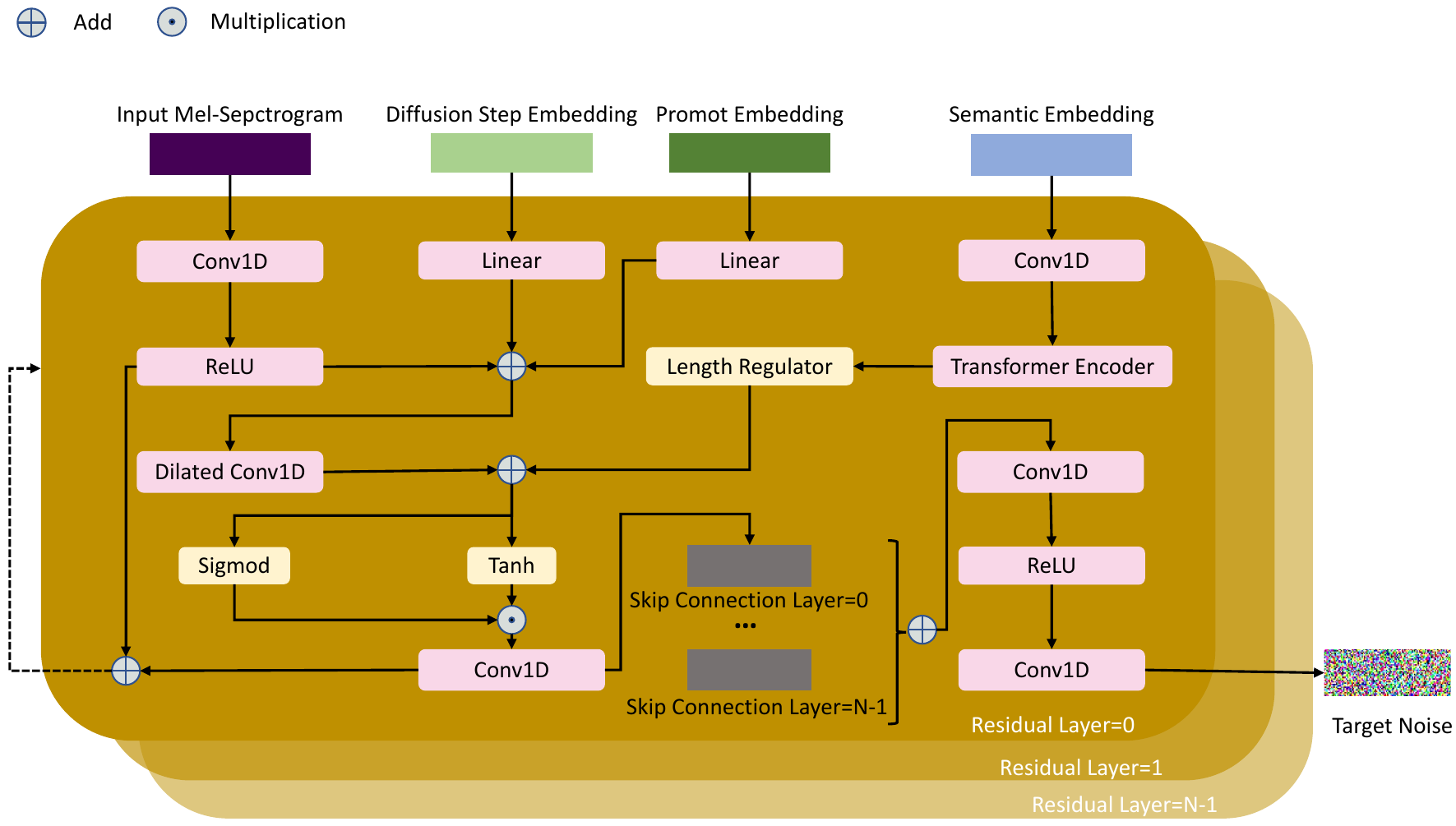}
 %\vspace{30pt}
 % \captionsetup{belowskip=-15pt}
 
 \caption{The architecture of acoustic diffusion model.}
 \label{fig:acoustic_model}
\end{figure}

\subsection{Conditional Diffusion Model}
\subsubsection{Diffusion  Formulation}
A diffusion model with ${T}$ diffusion steps consists of two processes: the diffusion process and the reverse process. The corresponding calculation for the acoustic diffusion model is shown in the Algorithms \ref{alg:training} and \ref{alg:sampling}. In the rest of this paper, the model uses $q(data), x_0, s, t$, and $p$ to represent data distribution, acoustic coding, semantic coding, diffusion step, and prompt embedding, respectively. One notable feature of the model is that it allows for closed-form sampling of $x_t$ at any timestep $t$ using $\bar{\alpha}_t$ and $\alpha_t$. The non-autoregressive network $\epsilon_{\theta}$ predicts $\epsilon$ from $x_t, t, p$, and $s$. The training objective is to minimize the unweighted variant of the ELBO\cite{ho2020denoising}, as shown in line 7 of Algorithm \ref{alg:training}. The sampling process is shown in Algorithm \ref{alg:sampling}, where $x_T \sim \mathcal{N}(0, I)$ is first sampled, followed by sampling $x_{t-1}\sim p_{\theta}(x_{t-1}|x_t)$ for $t=T, T-1,\cdots,1$. The output $x_0$ is the sampled data.

\subsubsection{Diffusion  Architecture}
As shown in Fig. \ref{fig:acoustic_model}, the acoustic diffusion model uses a bidirectional dilated convolution architecture with $N$ residual layers grouped into $m$ blocks, each containing $n = \frac{N}{m}$ layers. The dilation is doubled at each layer within each block. Skip connections from all residual layers are summed up, similar to WaveNet\cite{oord2016wavenet}. The model takes in semantic and prompt embeddings as conditional information. The semantic embedding is input to the transformer encoder, upsampled by the length regulator, and added as a bias term for the dilated convolution in each residual layer. The prompt embedding and diffusion step embedding are upsampled over the length and added to the input of each residual layer.

The other diffusion-based modules have similar structures but differ in input, diffusion-step, and conditional information. Fig. \ref{fig:proposed_model} shows that the duration diffusion model is conditioned on the phoneme sequence, while the semantic diffusion model is conditioned on the phoneme sequence upsampled by duration. The wave diffusion model is conditioned on the mel-spectrogram upsampled by frame length. The mel diffusion model is also conditioned on the phoneme sequence upsampled by duration and prompt embedding.

\section{Experiments}

\begin{figure}[t]
\begin{minipage}[t]{0.5\textwidth}
\begin{algorithm}[H]
  \caption{Training} \label{alg:training}
  \setlength{\belowdisplayskip}{50pt}
  \small
  \begin{algorithmic}[1]
    \Repeat
      \State $x_0, s \sim q(data)$
      \State $t \sim \mathrm{Uniform}(\{1, \dotsc, T\})$
      \State $p = {\hat{\mu}} + {\hat{\sigma}} \odot \phi ; \phi \sim \mathcal{N}(0, I)$
      \State $\varepsilon \sim \mathcal{N}(0, I)$
      \State $\bar{\alpha}_t = \prod_{i=1}^t\alpha_i$
      \State Take gradient descent step on
        \Statex $\nabla _\theta \left\| \varepsilon - \varepsilon_\theta((\sqrt{\bar\alpha_t} x_0 + \sqrt{1-\bar\alpha_t}\varepsilon), t, p, s) \right\|^2$
    \Until{converged}
  \end{algorithmic}
\end{algorithm}
\end{minipage}

\hfill
\begin{minipage}[t]{0.5\textwidth}
\begin{algorithm}[H]
  \caption{Sampling} \label{alg:sampling}
  \setlength{\belowdisplayskip}{50pt}
  \small
  \begin{algorithmic}[1]
    %\vspace{.04in}
    \State $x_T \sim \mathcal{N}(0, I)$
    \For{$t=T, \dotsc, 1$}
      
      \State $\mu_{\theta}(x_t, t, p, s) = \frac{1}{\sqrt{\alpha_t}}\left( x_t - \frac{1-\alpha_t}{\sqrt{1-\bar\alpha_t}} \varepsilon_\theta(x_t, t, p, s) \right)$
      \State $\sigma_{\theta}(x_t, t, p, s) =  \sqrt{\frac{1-\bar{\alpha}_{t-1}}{1-\bar{\alpha}_t}(1-\alpha_t)}$
      
      \State $x_{t-1} = \mu_{\theta} + \sigma_{\theta} \odot \psi; $ $\psi \sim \mathcal{N}(0, I)$ if $t > 1$, else $\psi = 0$
    \EndFor
    \State \textbf{return} $x_0$
    %\vspace{.04in}
  \end{algorithmic}
\end{algorithm}
\end{minipage}
\vspace{-1em}
\end{figure}

\subsection{Experimental Step}
 In our experiments, the semantic language model consists of 12 layer decoder-only transformer layers. The acousitc diffusion model has 30 residual layers, 64 residual channels, kernel size 3, dilation cycle $[1, 2, \cdots , 512]$, and the linear spaced schedule is $\beta_{t} \in [1\times10^{-4}, 0.05]$ ($T=200$). The other diffusion-based modules have similar structures but differ in diffusion step. The duration diffusion model is $T=5$. The semantic diffusion model is $T=200$. The wave diffusion model is $T=50$. The models are trained using 8 NVIDIA TESLA V100 32GB GPUs . Adam\cite{Kingma2014AdamAM} is used as the optimizer with an initial learning rate of 2e-4. 
 
\subsection{Compared Models and Datasets}
Due to the inability of standard (one-stage) TTS methods to perform minimally-supervised training, the test sets of {\bf Tacotron-VAE}\cite{qiang2023improving}, {\bf VALL-E}\cite{wang2023neural}, {\bf NaturalSpeech2}\cite{shen2023naturalspeech}, and {\bf Tri-Diff-Speech} include 3 hours of labeled data from each speaker. In contrast, the minimally-supervised TTS methods {\bf SpearTTS}\cite{zhang2023speak}, {\bf Diff-LM-Speech}, and {\bf Tetra-Diff-Speech} use test sets consisting of 15 minutes labeled data and 2.75 hours unlabeled data from each speaker. We combine an internal dataset with the AISHELL-3 dataset\cite{shi2020aishell}. All speech waveforms are sampled at 24kHz and converted to 40-band mel-spectrograms with a frame size of 960 and a hop size of 240. To ensure fairness, we modify all methods to utilize the same language model and diffusion model framework. Specifically, our proposed model's prompt encoder, wave diffusion model, and Grapheme-to-Phoneme (G2P) structure are identical across all compared models. The duration diffusion models of both {\bf Tri-Diff-Speech} (described in Sec \ref{sec:tri-diff}) and {\bf Tetra-Diff-Speech} (described in Sec \ref{sec:tetra-diff}) are also the same. Additionally, the semantic encoder and acoustic diffusion model of both {\bf Tetra-Diff-Speech} and {\bf Diff-LM-Speech} (described in Sec \ref{sec:diff-lm}) are identical. Hubert and Whisper's encoder are used as control groups for semantic coding. The front-end model structure is consistent with \cite{qiang2022back}.

\subsection{Test Metrics}
We conduct all subjective tests using 11 native judgers, with each metric consisting of 20 sentences per speaker. The test metrics used in the evaluation include prosody measurement, which involves mean square error for pitch ({\bf MSEP}) and duration ({\bf MSED}) to assess prosody similarity against ground-truth speech, word error rate ({\bf WER})(200 sentences per speaker), which utilizes an ASR model to transcribe the generated speech, and mean opinion score ({\bf MOS}), which verifies speech quality and similarity in expected speaking prosody and timbre between source speech and synthesized speech.

\begin{table}[]
% \captionsetup{skip=0pt} % 设置标题与表格之间的间距为10pt
\vspace{20pt}
 \caption{Prosody Measurement \& WER}
 \label{tab:prosody}
 \centering
 \resizebox{\linewidth}{!}{ 
\begin{tabular}{llll}
\hline
Model              & MSEP                    & MSED      & WER      \\ \hline
Tacotron-VAE\cite{qiang2023improving}          & 97.4          & \textbf{18.7} & 7.8          \\ \hline
VALL-E\cite{wang2023neural}                 & 98.6         & 19.5          & 6.1         \\ \hline
NaturalSpeech2\cite{shen2023naturalspeech} & 95.9          & 25.1          & \textbf{4.5} \\ \hline
SpearTTS(Hubert)\cite{zhang2023speak}      & 110.5         & 19.0          & 8.5          \\ \hline
Tetra-Diff-Speech(Hubert)                                   & 103.5         & 20.1          & 4.6          \\ \hline
Tetra-Diff-Speech(Whisper)                                  & 104.4         & 21.2          & 4.6          \\ \hline
Diff-LM-Speech(Hubert)                                      & 107.2         & \textbf{18.7}          & 7.2          \\ \hline
Tri-Diff-Speech                                             & \textbf{95.2} & 19.0          & \textbf{4.5} \\ \hline
\end{tabular}
}
\end{table}

\begin{table}[]
% \captionsetup{skip=0pt} % 设置标题与表格之间的间距为10pt
\vspace{20pt}
 \caption{Mean Opinion Score (MOS)}
 \label{tab:mos}
 \centering
\resizebox{\linewidth}{!}{ 
\begin{tabular}{llll}
\hline
Model                      & Prosody Sim           & Speaker Sim           & Speech Quality         \\ \hline
Tacotron-VAE               & 3.82 ± 0.072          & 3.92 ± 0.087          & \textbf{4.01 ± 0.023}          \\ \hline
VALL-E                     & 3.64 ± 0.050          & 3.70 ± 0.052          & 3.61 ± 0.013          \\ \hline
NaturalSpeech2             & 3.73 ± 0.054          & 4.04 ± 0.086          & 3.79 ± 0.070          \\ \hline
SpearTTS(Hubert)           & 3.60 ± 0.059          & 3.68 ± 0.030          & 3.50 ± 0.081          \\ \hline
Tetra-Diff-Speech(H)  & 3.89 ± 0.013          & 3.71 ± 0.077          & 3.83 ± 0.002          \\ \hline
Tetra-Diff-Speech(W) & 3.79 ± 0.098          & 3.93 ± 0.057          & 3.99 ± 0.017          \\ \hline
Diff-LM-Speech(H)     & 3.80 ± 0.090          & 3.71 ± 0.015          & 3.88 ± 0.042          \\ \hline
Tri-Diff-Speech            & \textbf{3.90 ± 0.047} & \textbf{4.06 ± 0.010} & \textbf{4.01 ± 0.080} \\ \hline
\end{tabular}
}
\end{table}

\subsection{Results}
The results in Table \ref{tab:prosody} demonstrate that one-stage models, including Tri-Diff-Speech, NaturalSpeech2, VALL-E, and Tacotron-VAE, outperform two-stage models like Spear-TTS, Tetra-Diff-Speech, and Diff-LM-Speech in terms of MSEP. Additionally, we observed that the semantic coding extracted by existing models suffers from excessive redundancy and dimension explosion, leading to difficulties in prediction from text and cumulative errors. Due to the limited number of open-source Mandarin datasets, we plan to further verify this conclusion in future work. In terms of MSED, Diff-LM-Speech achieves the best results, while Tri-Diff-Speech and Tetra-Diff-Speech also perform better than the non-autoregressive structure of NaturalSpeech2 due to the duration diffusion model. Furthermore, Tri-Diff-Speech and NaturalSpeech2 have significant advantages over other autoregressive structures (VALL-E, Diff-LM-Speech, etc.) in synthesizing robust speech due to their non-autoregressive structure, as demonstrated by the WER results.

Table \ref{tab:mos} reveals that the proposed methods outperform SpearTTS, NaturalSpeech2, and VALL-E in terms of prosody similarity MOS. This is due to the introduction of more randomness in the duration diffusion model, which enables diverse prosodic expressions. Among them, Tri-Diff-Speech achieves the best results. For speaker similarity MOS, both Tri-Diff-Speech and NaturalSpeech2 with non-autoregressive structures perform better. Moreover, all models using mel-spectrogram as acoustic features achieve better speech quality MOS scores than those with discrete acoustic coding. Specifically, Tri-Diff-Speech and Tacotron-VAE achieve the best results, highlighting the importance of continuous acoustic features for speech quality.

\section{Conclusions and future work}
In this paper, we propose three progressive methods, namely Diff-LM-Speech, Tetra-Diff-Speech, and Tri-Diff-Speech, to address several issues in existing systems. These issues include the high-frequency waveform distortion of discrete speech representations, the prosodic averaging problem caused by the duration prediction model, etc. The non-necessity of existing semantic coding models is verified. In future work, we expect to design an intermediate representation extraction method for minimally-supervised TTS.

\section{Acknowledgments}
This work is supported by the National Natural Sciencel Foundation of China (No. U23B2053, 62101553).

\vfill\pagebreak

\bibliographystyle{IEEEbib}

\footnotesize
\bibliography{strings,refs}

\end{document}